\documentstyle[aps,preprint]{revtex}
\begin{document}
\preprint{KSUCNR-002-95, MSUCL-1015}
\draft
\title{Non-equilibrium modifications of the nuclear equation of state}
\author{George Fai\thanks{
electronic mail: fai@ksuvxd.kent.edu}
}
\address{Institute for Nuclear Theory, University of Washington,
Seattle, WA 98195,
and \\
Center for Nuclear Research, Department of
Physics, Kent State University, Kent, OH 44242}
\author{Pawel Danielewicz\thanks{
electronic mail: danielewicz@nscl.nscl.msu.edu}
}
\address{Institute for Nuclear Theory, University of Washington,
Seattle, WA 98195,
and \\
National Superconducting Cyclotron Laboratory and Department of
Physics and Astronomy, Michigan State
University, East Lansing, MI 48824}

\maketitle
\begin{abstract}
Non-equilibrium modifications of the nuclear equation of state
are assessed in the framework of Extended Irreversible
Thermodynamics. The expected size of the
non-equilibrium corrections to quantities like the nuclear-matter
incompressibility $K$ and the liquid-vapor critical
temperature $T_c$ are compared to the uncertainties associated
at present with the usual equilibrium values.
The quantitative estimates are obtained in situations
characterized by
a stationary planar shear flow of realistic magnitude, or by a
stationary temperature gradient.
\end{abstract}

The study of nuclear collisions became a major subfield of nuclear physics
in the past two decades. Due to the complicated nature of the physics
and due to the many degrees of freedom involved, tools from
statistical physics
and thermodynamics proved to be efficient in modeling and analyzing
these collisions. For example, the concept of nuclear temperature
is commonly used. However, the applicability of equilibrium concepts
like temperature to the nuclear collision scenario
is frequently questioned.

It is hard to believe that equilibrium is reached during
the short lifetime of the collision system in a high-energy
nuclear collision, and thus the above question is well-justified.
One answer, which tries to avoid
the problem of equilibration in each collision, points out that these
experiments deal with very large samples of identically prepared
collision systems. What we have then, is an~ensemble of
events, while the usual introduction to statistical physics proceeds
via postulating the identity of time averages with ensemble
averages (ergodicity). In~nuclear collisions, we deal with
these ensembles directly. Each member of the
ensemble need not be in equilibrium, and the quantities used to characterize
the ensemble (e.g.~temperature) are ensemble averages. Due to
the relatively
small number of degrees of freedom (compared to e.g.~Avogadro's
number), the
distributions are in fact much wider than in conventional statistical
physics, leading to the important role of fluctuations in these
collisions.

Many people find the above argument unsatisfactory. In~this
Letter we
consider an~alternative point of view. We want to be able to
use
statistical-physics concepts like temperature without relying
on the assumption of equilibrium.
One may distinguish between global
equilibrium of the system (or a well-defined subsystem) and local
equilibrium (such as in the viscous hydrodynamic
model). Here we are particularly
interested in the situation when not even local equilibrium is established.
The~question then becomes: is it possible to
describe and usefully quantify the deviation from equilibrium? Similar
problems are addressed in non-equilibrium
thermodynamics\cite{proc83}.
In~particular, attempts to study non-equilibrium fluids
have been made in the framework of ``Extended Irreversible
Thermodynamics''~(EIT)\cite{mul67,cpg83}.

EIT starts with the assumption of a~local thermodynamic
potential (see~below),
which depends (in addition to its usual variables) on the dissipative fluxes
in the system such as the~heat flux or the viscous part of the
pressure (or~momentum-flux) tensor.
The dissipative fluxes are taken as additional independent variables.
While the evolution of the standard variables is governed by the conservation
equations of mass, energy and momentum, EIT derives further equations
for the dissipative fluxes. These equations can be used to calculate
non-equilibrium corrections to such quantities as the isothermal
nuclear-matter
incompressibility,~$K= 9/ (\rho \, \kappa_T)$, where $\kappa_T$ is
the compressibility usually defined as~$\kappa_T = \rho^{-1}
(\partial \rho / \partial p )_T$.  Although equilibrium in the
common sense is not assumed within EIT, systems are easiest
to discuss when stationary.
The~examples we
wish to
consider are those of a~planar flow under shear, and of
a~constant temperature gradient. The~nuclear collision scenario
is
of course far from stationary. We~believe, though, that
an~estimate
under these assumptions will indicate the order of magnitude of the
necessary non-equilibrium corrections.

The value of $K$ for nuclear matter has been subject to intense debate.
Values have been extracted from the giant monopole
resonance\cite{blaizot80,sharma88},
supernova calculations\cite{baron87}, and from
nuclear collisions assuming
momentum-independent~\cite{bertsch84,kruse85}
and momentum-dependent~\cite{gale87,csernai92,mishra93} mean
fields.
In~general, the~incompressibility (isothermal or adiabatic) is
expected to be a~function of density and
temperature. Numerical values quoted in the nuclear context normally
refer to standard density and zero temperature.
Recently, information from several independent experiments on
sideward flow was assessed within the framework of a~transport
model to arrive at 165~MeV $\leq K \leq$ 220~MeV\cite{pan93}.
We~will approximately represent this as $K = 190 \pm 30$~MeV
(see also\cite{zha94}).
We~assume that $\pm 30$ MeV is the characteristic uncertainty
associated with the equilibrium value of $K$ in a reasonable range
of densities and temperatures.
While the value of $K$ may be further discussed on the basis of different
calculations, to try to achieve better precision in similar studies
appears to require a very large effort.

In what follows we focus on the {\it uncertainty} in the
value of the isothermal incompressibility.
Note that the extracted values of $K$ refer to nuclear matter in
equilibrium, even though transport models do not rely on equilibrium.
It is interesting to compare the estimated EIT
non-equilibrium correction to the quoted uncertainty:
significant corrections would indicate that the
equilibrium value of $K$ is not really relevant in nuclear collisions.
If~the non-equilibrium corrections are comparable to the
uncertainty, then efforts to determine~$K$ with better
precision are not likely to be meaningful.

In order to estimate the non-equilibrium contributions to the
incompressibility or other quantities, specific situations
must be considered.  Within~EIT\cite{mul67,cpg83}, the complete
differential of the Helmholtz free energy per particle is
represented locally~as
\begin{equation}
d f = -s \, dT - p \, d \rho^{-1} + \alpha \, {\bf q} \cdot d
{\bf q} + \beta  \,
\overline{\overline{ \Pi}}_v : d
\overline{\overline{ \Pi}}_v \,\,\, ,
\label{df}
\end{equation}
where
$\overline{\overline{ \Pi}}_v$ is
a~traceless
viscous pressure tensor proportional to viscosity~$\eta$ and to
velocity derivatives, double dot indicates double contraction,
${\bf q} =
\kappa \nabla T$ is the heat flux, and $\kappa$ stands for heat conduction.
As~the~nuclear incompressibility can be
determined
from sideward flow in energetic (few hundred MeV/nucleon)
collisions,
where shear develops at the center of a~colliding
system\cite{dan95},
we~choose the~situation of a~planar shear, characterized~by
\begin{equation}
\overline{\overline{ \Pi}}_v = \left( \begin{array}{ccc}
                        0 & {\cal P} & 0 \\
                 {\cal P} &     0    & 0 \\
                        0 &     0    & 0   \end{array} \right)
\,\, ,
\label{Piv}
\end{equation}
with ${\cal P} = \eta \, (\partial u / \partial x)$, where $u$
is the velocity along the $y$ axis ($x$ and $y$ axes perpendicular to
the beam direction $z$), for assessing the corresponding
correction. The coefficient
in the shear term in~(\ref{df}) is
estimated~as\cite{cpg83}
\begin{equation}
\beta \simeq {\tau_v \over 2 \eta \rho} \,\,\, ,
\end{equation}
where $\tau_v$ is the relaxation time for~$\overline{\overline{
\Pi}}_v$.  For~a~rarefied gas, with~$\tau_v \approx (\rho
\sigma v)^{-1}$ and $\eta \simeq mv/3 \sigma$, the~coefficient
reduces~to
\begin{equation}
\beta \simeq {1 \over 2 \rho^2 T} \,\,\, ,
\label{beta}
\end{equation}
see also\cite{mul67}.  From the equality of mixed derivatives
of the Helmholtz energy, assuming an~inverse quadratic
dependence of~$\beta$ on density, the~pressure under shear
alone becomes
\begin{equation}
p = p_0 - \beta \, \rho \,
\overline{\overline{ \Pi}}_v :
\overline{\overline{ \Pi}}_v  = p_0 - 2 \, \beta \, \rho \,
{\cal P}^2 ,
\label{p=}
\end{equation}
where $p_0 = p_0 (\rho , T)$ is the pressure for given $\rho$ and
$T$ in the absence of shear.

Using (\ref{p=}) we find for the incompressibility (see
also\cite{perez83})
\begin{equation}
K = 9 \left( {\partial p \over \partial \rho} \right)_T
\simeq K_0 + 18 \, \beta \, {\cal P}^2 ,
\label{K=}
\end{equation}
where $K_0 = 9 (\partial p_0 / \partial \rho)_T$ is the equilibrium
value. Thus the incompressibility increases under shear.
The~larger
is the~density, cf.~(\ref{beta}), for a~given $\gamma =
\partial u / \partial x$, termed shear rate, the~lower is the
nonequilibrium correction, consistent with the~shortening of
the~mean free path.

We next assess the actual magnitude of the correction to the
incompressibility
under typical circumstances in collisions. For~this purpose we use
results of a~recent
transport simulation.  Figure~11 of~Ref.\cite{dan95} shows
density, velocity-component, and entropy profiles across
the~center of a~Au~+~Au system at 400~MeV/nucleon, at an
intermediate impact parameter, during the~compression stage.
The~shear rate at the center of the reaction in the figure
is $\gamma \approx 0.07 \, c$ and the density is $\rho \approx
0.30$ fm$^{-3}$.  Given the density and entropy there,
we~estimate a temperature of~$T \approx 44$ MeV. The nuclear
viscosity for these parameter values is about $\eta \approx
55$~MeV/fm$^2 c$\cite{dan84}.  Combining the~above results and
Eqs.~(\ref{beta})
and~(\ref{K=}), we~obtain $K - K_0 \sim 20$~MeV.  Similar
corrections are obtained at lower beam energies, as long as effects
of Fermi degeneracy do not yet enter the picture. We~conclude that
the nonequilibrium correction is comparable to the present
uncertainty associated with the determinations of the
incompressibility.

Another quantity of interest, the critical temperature of the
nuclear liquid-vapor phase
transition\cite{finn82,curtin83,campi88}
can be used as a further illustration
of the corrections to the equilibrium values obtained in EIT.
This~transition is explored in low-energy
reactions, in central
and/or~in spectator regions, where shear may be
low,
but~sizeable temperature differences are possible between the
matter
exposed~to and shielded from vacuum. The~outward heat
flow can modify local thermodynamic quantities and equations of
state in a manner similar to the case of shear.  Both shear
and heat flow are known to reduce critical temperatures in
macroscopic media\cite{jou83}.  For this illustration, we
adopt a~simple equation of state of the Van der Waals type
for nuclear matter in the transition region, and consider
a~constant temperature gradient.

The coefficient in the heat term in~(\ref{df}) may be estimated
as\cite{jou83}
\begin{equation}
\alpha \simeq {\tau_q \over \kappa \rho T} \,\,\, ,
\end{equation}
where $\tau_q$ is the relaxation time for ${\bf q}$, and, for a~rarefied
gas with~$\kappa \simeq c_p v/3 \sigma$,
\begin{equation}
\alpha \simeq {2 \over 5} \, {m \over \rho^2 \, T^2} \,\,\, ,
\label{alpha}
\end{equation}
see also\cite{mul67}.
Under an assumption of inverse quadratic dependence of~$\alpha$
on density, the pressure, for a~given heat flux,
becomes
\begin{equation}
p = p_0 - \alpha \, \rho \, q^2 ,
\end{equation}
where $p_0 = p_0 (\rho,T)$ is the equilibrium pressure.
Insertion of this result into the~Van der Waals equation
of state then yields
\begin{equation}
\left( p + { a \over {\sl v}^2} + {\alpha \,  q^2 \over {\sl
v}} \right) \, \big( {\sl v} - b \big) = T \, ,
\label{gvdW}
\end{equation}
where ${\sl v} = 1/\rho$ is the specific volume.
The critical-point conditions
\begin{equation}
\left(\frac{\partial p}{\partial {\sl v}}\right)_{T,{\bf q}} = 0
\hspace{3em} \mbox{ and}\hspace{3em}
\left(\frac{\partial^2 p}{\partial {\sl v}^2}\right)_{T,{\bf q}} = 0 \,\,\, ,
\label{crit}
\end{equation}
give then a~lowering of the critical temperature\cite{jou83}.
To leading order in~$q^2$, the~temperature~is
\begin{equation}
T_c (q) = T_c^0 - \frac{4}{9} \, \alpha \, q^2  \,\, .
\label{tcrit}
\end{equation}

The~phase-transition region is crossed in reactions when matter
expands to the vacuum and nuclear fragments are formed.
If the expansion is isentropic and statistical effects are ignored,
then the product $\rho \, T^{-3/2}$ may be considered constant. Further,
if the density changes from  $\rho \, {\buildrel < \over
\sim} \, \rho_0$ in the interior of the expanding matter
to $\rho \, {\buildrel > \over
\sim} \, 0.2 \, \rho_0$ at an~outside freeze-out surface, over
a~characteristic distance $\sim (5-10)$~fm, then the~resulting
temperature gradients are of the order of 1~MeV/fm in the
critical region $T_c \sim 15$~MeV.  Given that actual
gradients would be somewhat moderated by the heat conduction (as~is
supported by numerical simulations\cite{dan95}), we~expect heat
fluxes\cite{dan84} of magnitude $q \, {\buildrel < \over
\sim} \, 0.1$~MeV$\, c$/fm$^3$.  With~(\ref{alpha}) and $\rho \sim
0.5 \, \rho_0$ in the critical region, we~finally estimate
the~correction to the critical temperature to be of the order
of~${\buildrel < \over \sim} \, 2$~MeV.

On examining such corrections as in~(\ref{tcrit}) or~(\ref{p=})
for a~rarefied gas, it~is seen that these corrections
are associated with the variation of temperature or of
collective
velocity over the~distance of one mean free path.  It~is
plausible that the critical temperature of uncharged infinite
matter could be less relevant for fragment formation, than
e.g.~the limiting temperature\cite{lev85,sur87}, the~maximum
temperature a~residual nucleus could withstand given
Coulomb interactions and the~lowering of surface tension.
In~any case, EIT gives a~typical magnitude of corrections on
account of dissipative fluxes, and these corrections are
comparable to typical uncertainties associated with the choice
of interactions\cite{bal94}.

To summarize,
in this Letter we quantified the deviations from equilibrium
expected
in intermediate energy nuclear collisions. To achieve this goal, we
used the tools of Extended Irreversible Thermodynamics.
As specific examples,
the non-equilibrium corrections to the incompressibility and
to the critical temperature of the liquid-vapor phase transition were
examined.  Our numerical estimates are based on a stationary
planar shear
flow and on a constant temperature gradient, respectively.
The~comparison of the non-equilibrium modifications to the
uncertainties associated with equilibrium values
reveals that these are comparable to each other.
This casts doubt on the reducibility of
the uncertainties on the basis of nuclear collision data.
The nonequilibrium modifications imply
that nuclear matter acts effectively stiffer and
displays a lower critical temperature than
in equilibrium.

\acknowledgements

We are grateful for the hospitality extended to us at
the Institute for Nuclear Theory at the University of
Washington, where the first part of this work was completed.
Useful comments by M.~Lisa are acknowledged.
This work was partially
supported by the Department of
Energy under Grant Nos.\ DOE/DE-FG02-86ER-40251 and
DOE/DE-FG06-90ER40561 and by
the National Science Foundation under Grant No.~PHY-9403666.

\end{document}